\documentstyle[11pt,fleqn]{article}
\title{Classification of Seven-Vertex Solutions of The Coloured Yang-Baxter Equation}
\author{Shi-Kun Wang , Hai-Tang Yang\\
Institute of Applied Mathematics\\
Academia Sinica, Beijing, 100080\\
People's Republic of China\\
Ke Wu\\
Institute of Theoretical Physics\\
Academia Sinica, Beijing, 100080\\
People's Repulic of China}
\def\v4{vskip4mm}
\begin{document}
\maketitle
\begin{minipage}[t]{4in}
Abtract. In this paper all seven-vertex type solutions of the coloured
Yang-Baxter equation dependent on spectral as well as coloured parameters are
given. It is proved that they are composed of five groups of basic solutions, two groups of their degenerate forms
up to five solution transformations. Moreover, all solutions can be claasified
into two types called Baxter type and free-fermion type.
\end{minipage}

\section{Introduction}

The Yang-Baxter equation (YBE), first appeared in refs. [1-3] plays a prominent
role in many branches of physics. In the field theories, the YBE can completely determine
 the S-matrix for the two-body scattering amplitudes in the
multi-particle scattering processes [4]. In two dimensional integrable models
of quantum fields theories and 
statistical models, the YBE provides an essential and consistency condition
in establishing the integrability and solving the models [5-7]. In
conformal field theory, the YBE is an important equation related to the KZB
equation [8]. Motivated by the important role of the YBE, much attention has
been directed to the search for the solutions of YBE [9-13].

From the YBE's birth, there have several alterations of it. In the beginning,
for simplifying the solving process, people first try to deal with the
spectral independent YBE (simple YBE) and classical YBE (without quantum
parameter) and have achieved great success. Later, some generalizations
were given. The coloured YBE (CYBE) which depends on spectral as well as
coloured parameter is one of them. Since CYBE  concerns the free-fermion
model in a magnetic field, multi-variable invariants of link and representations
of quantum algebras and so on [12-16]
It has also attracted a lot of research to find the solutions of CYBE [14-23].
The first solution of CYBE with six-vertex can be found in ref. [14].
The eight-vertex solution of the CYBE has been investigated previously in
free fermion model in magnetic field by Fan C. and Wu F. Y. [15], who first
introduce a simple relation between the weight functions in the model,
so-called  free-fermion condition. Based on the work and the condition,
V. V. Bazhanov and Yu. G. Stroganov obtained a eight-vertex solution of
the CYBE [19]. J. Murakami gave another eight-vertex solution in discussing
multi-variable invariants of links. The first and third authors in this
paper gave all six-vertex and eight-vertex types solutions of CYBE and
classified them into two kinds called Baxter and Free-Fermion types [22,23].

The main theme of this paper is to give and classify seven-vertex solutions of CYBE by
a computer algebraic method. Moreover, it is
proved from a theorem in ref. [24] that all seven-vertex type solutions of
CYBE have indeed obtained in this paper.
In section 2 we will introduce the
symmetries or solution transformation for CYBE, then by the symmetries and
a computer algebraic method to find the most simple system of equations,
including differential equation, the solution set of which covers the one
of CYBE. In section 3 and section 4 we will seperatly give all solutions of the seven-vertex CYBE and classify them into two
kinds called Baxter and Free-Fermion types.
\vskip7mm

\section{The symmetries and initial conditions of the coloured Yang-Baxter equation}

The coloured Yang-Baxter equation means the following matrix equation:
$$\begin{array}{l}
\check R_{12}(u,\xi,\eta)\check R_{23}(u+v,\xi,\lambda)\check R_{12}(v,\eta,\lambda)=\check R_{23}(v,\eta,\lambda)\check R_{12}(u+v,\xi,\lambda)\check R_{23}(u,\xi,\eta)\\[4mm]
\check R_{12}(u,\xi,\eta)=\check R(u,\xi,\eta)\otimes E\\[4mm]
\check R_{23}(u,\xi,\eta)=E\otimes \check R(u,\xi,\eta)
\end{array}
\eqno{(2.1)}$$
where $\check R(u,\xi,\eta)$ is a matrix function of $N^2$ dimension of
$u$,$\xi$ and $\eta$. $E$ is the unit matrix of order $N$ and $\otimes$
means the tensor product of two matrices. $u,v$ are spectral parameters
and $\xi,\eta$ are colour parameters. If we let $\xi=\eta=0$, we can get
the usual Yang-Baxter equation. Siminlarly, if we have $u=v=0$, then (2.1)
will be reduced to the pure coloured Yang-Bater equation.

In this paper, the main interst for the equation (2.1) is to discuss the
solutions of the senven-vertex-type as following:
$$\check R(u,\xi,\eta)=\left(
\begin{array}{cccc}
R_{11}^{11}(u,\xi,\eta)&0&0&R_{22}^{11}(u,\xi,\eta)\\
0&R_{12}^{12}(u,\xi,\eta)&R_{21}^{12}(u,\xi,\eta)&0\\
0&R_{12}^{21}(u,\xi,\eta)&R_{21}^{21}(u,\xi,\eta)&0\\
0&0&0&R_{22}^{22}(u,\xi,\eta)
\end{array}
\right)
\eqno{(2.2)}$$
The seven weight functions in (2.2) are denoted by
$$\begin{array}{ll}
a_1(u,\xi,\eta)=R_{11}^{11}(u,\xi,\eta)&a_5(u,\xi,\eta)=R_{12}^{21}(u,\xi,\eta)\\[4mm]
a_2(u,\xi,\eta)=R_{12}^{12}(u,\xi,\eta)&a_6(u,\xi,\eta)=R_{21}^{12}(u,\xi,\eta)\\[4mm]
a_3(u,\xi,\eta)=R_{21}^{21}(u,\xi,\eta)&a_7(u,\xi,\eta)=R_{11}^{22}(u,\xi,\eta)\\[4mm]
a_4(u,\xi,\eta)=R_{22}^{22}(u,\xi,\eta)&
\end{array}$$
In this paper, we only consider $a_i(u,\xi,\eta)(i=1,2,\ldots,7)$ are meromorphic functions of $u,\xi,\eta$ and $a_i(u,\xi,\eta)\not\equiv 0$
which means the nondegenerate solutions. Throughout this paper, we let
$$\begin{array}{llll}
u_i=a_i(u,\xi,\eta)&v_i=a_i(v,\eta,\lambda)&w_i=a_i(u+v,\xi,\lambda)&i=1,2,\ldots ,8
\end{array}$$
For the seven-vertex-type solutions, the matrix equation (2.1) is equivalent
to the following 19 equations:
$$
\begin{array}{l}
u_2w_3v_2-u_3w_2v_3=0
\end{array}
\eqno{(2.3a)}$$
$$\left.
\begin{array}{l}
u_1w_5v_2-u_5w_1v_2-u_3w_2v_5=0\\
u_2w_6v_1-u_2w_1v_6-u_6w_2v_3=0\\
u_1w_2v_1-u_2w_1v_2-u_6w_2v_5=0
\end{array}
\right\}
\eqno{(2.3b)}$$
$$\left.
\begin{array}{l}
u_4w_6v_2-u_6w_4v_2-u_3w_2v_6=0\\
u_2w_5v_4-u_2w_4v_5-u_5w_2v_3=0\\
u_4w_2v_4-u_2w_4v_2-u_5w_2v_6=0
\end{array}
\right\}
\eqno{(2.3c)}$$
$$\left.
\begin{array}{l}
u_1w_5v_3-u_5w_1v_3-u_2w_3v_5=0\\
u_3w_6v_1-u_3w_1v_6-u_6w_3v_2=0\\
u_1w_3v_1-u_3w_1v_3-u_6w_3v_5=0\\
u_1w_1v_7+u_7w_2v_4-u_5w_5v_7-u_2w_7v_1=0\\
u_1w_7v_5+u_7w_6v_2-u_3w_5v_7-u_6w_7v_1=0\\
u_1w_7v_3+u_7w_6v_6-u_7w_1v_1-u_4w_3v_7=0
\end{array}
\right\}
\eqno{(2.3d)}$$
$$\left.
\begin{array}{l}
u_4w_6v_3-u_6w_4v_3-u_2w_3v_6=0\\
u_3w_5v_4-u_3w_4v_5-u_5w_3v_2=0\\
u_4w_3v_4-u_3w_4v_3-u_5w_3v_6=0\\
u_4w_4v_7+u_7w_2v_1-u_6w_6v_7-u_2w_7v_4=0\\
u_4w_7v_6+u_7w_5v_2-u_3w_6v_7-u_5w_7v_4=0\\
u_4w_7v_3+u_7w_5v_5-u_7w_4v_4-u_1w_3v_7=0
\end{array}
\right\}
\eqno{(2.3e)}$$
Assuming $\check R(u,\xi,\eta)$ is the solution of equation (2.1). We can find there are five symmetries in the system of equations (2.3):
\begin{description}
\item[(A) Symmetry of interchanging indices.] The system of equations
(2.3) is invariant if we interchang the two sub-indices 1 and 4 as well
as the two sub-indices 5 and 6.
\item[(B) The scaling symmetry.] Multiplication of the solution $\check
R(u,\xi,\eta)$ by an arbitary function ${\cal F}(u,\xi,\eta)$ is still
a solution of the equation (2.1).
\item[(C) Symmetry of the weight functions.] If the weight functions
$a_2(u,\xi,\eta)$ , $a_3(u,\xi,\eta)$ , $a_7(u,\xi,\eta)$ are replaced by the new
weight functions:
$$\begin{array}{ll}
\tilde a_2(u,\xi,\eta)=\frac{N(\xi)}{N(\eta)}a_2(u,\xi,\eta)&\tilde a_3(u,\xi,\eta)=\frac{N(\eta)}{N(\xi)}a_3(u,\xi,\eta)\\[4mm]
\tilde a_7(u,\xi,\eta)=\frac{1}{sN(\xi)N(\eta)}a_7(u,\xi,\eta)&
\end{array}$$
respectively, or $a_7(u,\xi,\eta),a_5(u,\xi,\eta)$ and $a_6(u,\xi,\eta)$ are replaced by $-a_7(u,\xi,\eta), -a_5(u,\xi,eta)$
and $-a_6(u,\xi,\eta)$, where $N(\xi)$ is an arbitary function of $\xi$ and
$s$ is a complex constant, the new matrix $\check R(u,\xi,\eta)$ is still a solution of (2.1).
\item[(D) Symmetry of spectral parameters.] If we take the new spectral
parameter $\tilde u=\mu u$ where $\mu$ is a complex constant,the new matrix $\check R(\tilde u,\xi,\eta)$ is still a solution of (2.1).
\item[(E) Symmerty of the coloured parameters.] If we take the new coloured
parameters $\zeta=f(\xi)$, $\theta=f(\eta)$, where $f(\xi)$ is an arbitary function, then the new matrix $\check R(u,\zeta,\theta)$ is also a solution of the equation (2.1).
\end{description}

The five symmetries (A)-(E) are called solution transformations {\bf A-E} of seven-vertex-type solutions of coloured Yang-Baxter equation (2.1), respectively.

Dividing both sides of equaiton (2.3a) by $u_2w_2v_2$, we get
$$f(u+v,\xi,\lambda)=f(u,\xi,\eta)f(v,\eta,\lambda)
\eqno{(2.4)}$$
where $f(u,\xi,\eta)=\frac{u_3}{u_2}$. Taking $u=v=\eta=0$ in (2.4) we get
$$f(0,\xi,\lambda)=f(0,\xi,0)f(0,0,\lambda)$$
Taking $u=v=\xi=0$ in (2.4) we have
$$f(0,0,\lambda)=f(0,0,\eta)f(0,\eta,\lambda)$$
Then we have 
$$f(0,0,\lambda)=f(0,0,\eta)f(0,\eta,0)f(0,0,\lambda)$$
This means
$$f(0,0,\eta)f(0,\eta,0)=1$$
Otherwise, it is easy to show that $f(u,\xi,\eta)=0$, i.e.$a_3(u,\xi,\eta)=0$.
Therefore, we have
$$f(0,\xi,\eta)=\frac{M(\xi)}{M(\eta)}
\eqno{(2.5)}$$
where $M(\xi)=f(0,\xi,0)$. On the other hand, if we differentiate both sides
of (2.4) with respect to the spectral variable $v$ and then set $v=0$,
$\lambda=\eta$, then
$$f^{'}(u,\xi,\eta)=f(u,\xi,\eta)f^{'}(0,\eta,\eta)$$
holds, where the dot means derivative to u and the simple formula
$$\left.
\frac{dH(u+v)}{dv}\right|_{v=0}=\frac{dH(u)}{du}$$
for any function $H(u)$ is used. Similarly, if we differentiate (2.4) with
respect to $u$ and then set $u=0$ and $\eta=\xi$ we have
$$f^{'}(v,\xi,\lambda)=f(v,\xi,\lambda)f^{'}(0,\xi,xi)$$
The two fomulae above imply $f^{'}(0,\xi,\xi)$ is a constant independent of
coloured parameter $\xi$. Hence
$$f(u,\xi,\eta)=\frac{M(\xi)}{M(\eta)}exp(ku)
\eqno{(2.6)}$$
where $k$ is a complex constant.

So up to the solution transformation {\bf B} and {\bf C}, we can assume
$$\begin{array}{ll}
a_2(u,\xi,\eta)=1&a_3(u,\xi,\eta)=exp(ku)
\end{array}$$
without losing generality. Then the system equation (2.3) can be simplified
to the following 12 equations:
$$\begin{array}{l}
u_1w_5-u_5w_1-v_5exp(ku)=0\\[4mm]
w_6v_1-w_1v_6-u_6exp(kv)=0\\[4mm]
u_1v_1-w_1-u_6v_5=0\\[4mm]
u_1w_1v_7+u_7v_4-u_5w_5v_7-w_7v_1=0\\[4mm]
u_1w_7v_5+u_7w_6-u_6w_7v_1-w_5v_7exp(ku)=0\\[4mm]
u_7w_6v_6-u_7w_1v_1+u_1w_7exp(kv)-u_4v_7exp(ku+kv)=0
\end{array}
\eqno{(2.7)}$$
plus six equation, which are called the counterparts of (2.7), obtained by interchanging the sub-indices 1 and 4 as well as 5 and 6 in each of the equation in (2.7).

Now we solve the equations obtained by setting u=0 and $\eta=\xi$ in (2.7).  It is easy to get:
$$\begin{array}{l}
a_1(0,\xi,\xi)=a_4(0,\xi,\xi)=1\\[4mm]
a_5(0,\xi,\xi)=a_6(0,\xi,\xi)=a_7(0,\xi,\xi)=0
\end{array}
\eqno{(2.8)}$$
which are called the initial conditions of (2.1) in this paper. Substituting (2.8) into 2.7 after letting $v=-u,\lambda=\xi$ we obtain
$$\begin{array}{l}
a_5(u,\xi,\eta)=-a_5(-u,\eta,\xi)exp(ku)\\[4mm]
a_6(u,\xi,\eta)=-a_6(-u,\eta,\xi)exp(ku)\\[4mm]
a_4(u,\xi,\eta)a_4(-u,\eta,\xi)=a_1(u,\xi,\eta)a_1(-u,\eta,\xi)\\[4mm]
a_7(u,\xi,\eta)a_1(-u,\eta,\xi)=-a_7(-u,\eta,\xi)a_4(u,\xi,\eta)
\end{array}
\eqno{(2.9)}$$

Differentiating both sides of all equations in (2.7) and their counterparts
with respect to the variable $v$ and letting $v=0, \lambda=\eta$, by the virtue of initial conditions (2.8) we can get:
$$\begin{array}{l}
u_1u_5^{'}-u_5u_1^{'}-m_5(\eta)exp(ku)=0\\[4mm]
u_6^{'}+u_6m_1(\eta)-u_1m_6(\eta)-ku_6=0\\[4mm]
u_1m_1(\eta)-u_1^{'}-u_6m_5(\eta)=0\\[4mm]
u_1u_7m_5(\eta)+u_7u_6^{'}-u_5m_7(\eta)exp(ku)-u_6u_7m_1(\eta)-u_6u_7^{'}=0\\[4mm]
ku_1u_7+u_1u_7^{'}+u_6u_7m_6(\eta)-u_1u_7m_1(\eta)-u_7u_1^{'}-u_4m_7(\eta)exp(ku)=0\\[4mm]
(u_1^2-u_5^2)m_7(\eta)+(m_4(\eta)-m_1(\eta))u_7-u_7^{'}=0
\end{array}
\eqno{(2.10a)}
$$
and their counterparts, where and throughout this paper unless obvious comment we denote
$$\begin{array}{ll}
a_i^{'}(u,\xi,\eta)=\frac{\partial a_i(u,\xi,\eta)}{\partial u}&\left.  m_i(\xi)=a_i^{'}(u,\xi,\eta)\right|_{(u=0,\eta=\xi)}\\
i=1,2,\ldots,7&
\end{array}$$
We call $m_i(\xi)$ Hamiltonian coefficients of weight functions with respect to the spectral parameter or simply cofficients. Sometimes we write $m_i$ instead of $m_i(\xi)$ for brevity.

If we differentiate (2.7) with respect to $u$ and let $u=0, \eta=\xi$ and then replace the variables $v$ and $\lambda$ by $u$ and $\eta$, we can get:
$$\begin{array}{l}
u_5m_1(\xi)+u_5^{'}-u_1m_5(\xi)-ku_5=0\\[4mm]
u_1u_6^{'}-u_1^{'}u_6-m_6(\xi)exp(ku)=0\\[4mm]
u_1m_1(\xi)-u_1^{'}-u_5m_6(\xi)=0\\[4mm]
u_1u_7m_1(\xi)-u_1u_7^{'}+u_1^{'}u_7+u_4m_7(\xi)-u_5u_7m_5(\xi)=0\\[4mm]
u_5u_7m_1(\xi)+u_5u_7^{'}-u_5^{'}u_7+u_6m_8(\xi)-ku_5u_7-u_1u_7m_6(\xi)=0\\[4mm]
(m_1(\xi)-m_4(\xi))u_7exp(ku)+(u_6^2-u_1^2)m_7(\xi)+u_7^{'}exp(ku)-ku_7exp(ku)=0
\end{array}
\eqno{(2.10b)}$$
and their counterparts.

If we differentiate (2.7) with respect to $\lambda$ and then let $v=0, \lambda=\eta$, we can get:
$$\begin{array}{l}
u_1u_5^{'}-u_5u_1^{'}-\check m_5(\eta)exp(ku)=0\\[4mm]
u_6^{'}+u_6\check m_1(\eta)-u_1\check m_6(\eta)=0\\[4mm]
u_1\check m_1(\eta)-u_1^{'}-u_6\check m_5(\eta)=0\\[4mm]
u_1u_7\check m_5(\eta)+u_7u_6^{'}-u_5\check m_7(\eta)exp(ku)-u_6u_7\check m_1(\eta)-u_6u_7^{'}=0\\[4mm]
u_1u_7^{'}+u_6u_7\check m_6(\eta)-u_1u_7\check m_1(\eta)-u_7u_1^{'}-u_4\check m_7(\eta)exp(ku)=0\\[4mm]
(u_1^2-u_5^2)\check m_7(\eta)+(\check m_4(\eta)-\check m_1(\eta))u_7-u_7^{'}=0
\end{array}
\eqno{(2.10c)}$$
and their counterparts.
Here we denote:
$$\begin{array}{ll}
u_i^{'}=\frac{\partial a_i(u,\xi,\eta)}{\partial \eta}&\left. \check m_i(\eta)=\frac{\partial a_i(v,\eta,\lambda)}{\partial \lambda}\right|_{(v=0,\lambda=\eta)}
\end{array}$$

Similarily, if we differentiate (2.7) with respect to $\xi$ and let $u=0,\eta=\xi$ and then replace $v,\lambda$ by $u,\eta$ we have:
$$\begin{array}{l}
u_5\hat m_1(\xi)+u_5^{'}-u_1\hat m_5(\xi)=0\\[4mm]
u_1u_6^{'}-u_1^{'}u_6-\hat m_6(\xi)exp(ku)=0\\[4mm]
u_1\hat m_1(\xi)-u_1^{'}-u_5\hat m_6(\xi)=0\\[4mm]
u_1u_7\hat m_1(\xi)-u_1u_7^{'}+u_1^{'}u_7+u_4\hat m_7(\xi)-u_5u_7\hat m_5(\xi)=0\\[4mm]
u_5u_7\hat m_1(\xi)+u_5u_7^{'}-u_5^{'}u_7+u_6\hat m_8(\xi)-u_1u_7\hat m_6(\xi)=0\\[4mm]
(\hat m_1(\xi)-\hat m_4(\xi))u_7exp(ku)+(u_6^2-u_1^2)\hat m_7(\xi)+u_7^{'}exp(ku)=0
\end{array}
\eqno{(2.10d)}$$
and their counterparts.
Here we denote:
$$\begin{array}{ll}
u_i^{'}=\frac{\partial a_i(u,\xi,\eta)}{\partial \xi}&\left.\hat m_i(\xi)=\frac{\partial a_i(u,\xi,\eta)}{\partial \xi}\right|_{(u=0,\eta=\xi)}
\end{array}$$
From the last equation of (2.10a) and its counterpart we have:
$$2u_7{'}=m_7(\eta)(u_1^2+u_4^2-u_5^2-u_6^2)
\eqno{(2.11a)}$$
We also get from the last equation of (2.10b) and its counterpart the following equation:
$$2u_7^{'}=m_7(\xi)(u_1^2+u_4^2-u_5^2-u_6^2)exp(-ku)+2ku_7
\eqno{(2.11b)}$$

Now there are two cases, $k=0$ and $k\not=0$. We will discuss them seperately in the later sections.

\section{The solutions of equation (2.1) in the case of $k\not=0$ }

Now, we first consider the case of $k\not=0$. From equations (2.11), we can obtain
$$\begin{array}{lll}
u_7^{'}\not=0&m_7(\xi)\not=0&m_7(\eta)\not=0
\end{array}
\eqno{(3.1)}$$
and we also have:
$$\frac{u_7^{'}-ku_7}{u_7^{'}}exp(ku)=\frac{ m_7(\xi)}{m_7(\eta)}$$
by solving the above equation we can get:
$$\begin{array}{ll}
a_7(u,\xi,\eta)=F_7(\xi,\eta)(exp(ku)-\displaystyle\frac{F_7(\xi,\xi)}{F_7(\eta,\eta)}&m_7(\eta)=kF_7(\eta,\eta)
\end{array}
\eqno{(3.2)}$$

The following work is to eliminate the five weight functions$\{w_1,w_4,w_5,w_6,w_7\}$ in the equation (2.7) and their counterparts. Then we get seven polynomial equations without weight functions $w_i(i=1,4,\ldots,7)$. If we differentiate the obtained seven equations with respect to the spectral parameter $v$, and let $v=0,\lambda=\eta$ and substitute the initial conditons (2.8) into them, we obtain the following seven polynomial equatons:
$$\begin{array}{l}
m_6(\eta)exp(ku)+(m_1(\eta)+m_4(\eta))u_4u_6-m_6(\eta)(u_1u_4+u_5u_6)-ku_4u_6=0\\[4mm]
m_5(\eta)exp(ku)+(m_4(\eta)+m_1(\eta))u_1u_5-m_5(\eta)(u_1u_4+u_5u_6)-ku_1u_5=0\\[4mm]
m_7(\eta)(u_1^2-u_4^2-u_5^2+u_6^2)+2u_7(m_4(\eta)-m_1(\eta))=0\\[4mm]
m_7(\eta)(u_1^2u_6-u_5^2u_6+u_5exp(ku))-(m_5(\eta)+m_6(\eta))u_1u_7-(k-m_1(\eta)-m_4(\eta))u_6u_7=0\\[4mm]
m_7(\eta)(u_1u_5^3-u_1^3u_5-u_1u_6exp(ku))+m_5(\eta)u_7exp(ku)+2(m_1(\eta)-m_4(\eta)u_5u_7+m_6(\eta)u_1u_4u_7-m_5(\eta)u_5u_6u_7=0\\[4mm]
m_7(\eta)(u_1^3-u_1u_5^2-u_4exp(ku))+u_1u_7(k+m_4(\eta)-3m_1(\eta))+(m_5(\eta)+m_6(\eta))u_6u_7=0\\[4mm]
m_7(\eta)u_4(u_1^2-u_5^2)+u_4u_7(k-m_1(\eta)-m_4(\eta))+(m_5(\eta)+m_6(\eta))u_5u_7=0[4mm]
\end{array}
\eqno{(3.3)}$$
As the second step, we eliminate $m_1,m_4$ and $m_5$ to get two systems of equations, which are equivalent to (3.3). The first is
$$\begin{array}{l}
m_5exp(ku)+(m_1+m_4)u_1u_5-m_5(u_1u_4+u_5u_6)-ku_1u_5=0\\[4mm]
m_7u_1u_5(u_4^2-u_1^2+u_5^2-u_6^2)+2m_5u_7(u_1u_4+u_5u_6-exp(ku))+(2k-4m_4)u_1u_5u_7=0\\[4mm]
m_7u_1(u_4^2u_5-u_5^2u_6+u_6exp(ku))-u_7(m_5exp(ku)+m_6u_1u_4-m_5u_5u_6)=0
\end{array}
\eqno{(3.4)}$$
which contains $m_1,m_4$ and $m_5$. The second is
$$\begin{array}{l}
(u_1u_4+u_5u_6-exp(ku))(-m_7u_4u_6(u_4^2u_5-u_5u_6^2+u_6exp(ku))+m_6u_7(u_4^2u_6-u_5^2u_6+u_5exp(ku)))=0\\[4mm]
(u_1u_4+u_5u_6-exp(ku))(m_7u_1u_5(u_6^2-u_4^2)+m_7exp(ku)(u_4u_5-u_1u_6)+m_6u_7(u_1u_4-u_5u_6))=0\\[4mm]
(u_1u_4+u_5u_6-exp(ku))(m_7u_5u_6(u_4u_6-u_1u_5)+m_7u_4u_5(u_5^2-u_4^2)+m_7exp(ku)(u_1u_5-u_4u_6)+m_6u_7(u_4^2-u_5^2))=0\\[4mm]
(u_1u_4+u_5u_6-exp(ku))(m_7u_4u_5(u_1u_5-u_4u_6)+m_7(u_5u_6-exp(ku))(u_6^2-u_5^2)+m_6u_7(u_4u_6-u_1u_5))=0
\end{array}
\eqno{(3.5)}$$
which do not contain $m_1,m_4$ and $m_5$. So
$$u_1u_4+u_5u_6=exp(ku)
\eqno{(3.6)}$$
or
$$\begin{array}{l}
-m_7u_4u_6(u_4^2u_5-u_5u_6^2+u_6exp(ku))+m_6u_7(u_4^2u_6-u_5^2u_6+u_5exp(ku))=0\\[4mm]
m_7u_1u_5(u_6^2-u_4^2)+m_7exp(ku)(u_4u_5-u_1u_6)+m_6u_7(u_1u_4-u_5u_6)=0\\[4mm]
m_7u_4u_5(u_1u_5-u_4u_6)+m_7(u_5u_6-exp(ku))(u_6^2-u_5^2)+m_6u_7(u_4u_6-u_1u_5)=0\\[4mm]
m_7u_5u_6(u_4u_6-u_1u_5)+m_7u_4u_5(u_5^2-u_4^2)+m_7exp(ku)(u_1u_5-u_4u_6)+m_6u_7(u_4^2-u_5^2)=0
\end{array}
\eqno{(3.7)}$$
will hold. In the third step, applying the fourth equation in (3.7) as a main equation to kill $m_6$ in the three other equations in (3.7) and then performing factorization of the new polynomial equations after killing $m_6$, we can obtain
$$\begin{array}{l}
m_7u_5u_6(u_4u_6-u_1u_5)+m_7u_4u_5(u_5^2-u_4^2)+m_7exp(ku)(u_1u_5-u_4u_6)+m_6u_7(u_4^2-u_5^2)=0\\[4mm]
m_7(u_5u_6-exp(ku))(u_1^2u_5-2u_1u_4u_6+u_4^2u_5-u_5^3+u_5u_6^2)u_5=0\\[4mm]
m_7(u_5u_6-exp(ku))(-u_1u_4^2u_6+u_1u_5^2u_6+u_4^3u_5-u_4u_5^3-u_1u_5exp(ku)+u_4u_6exp(ku))u_5=0\\[4mm]
m_7(u_5u_6-exp(ku))(-u_1^2u_4+2u_1u_5u_6+u_4^3-u_4u_5^2-u_4u_6^2)u_5=0
\end{array}
\eqno{(3.8)}$$
which are equivalent to (3.7). We have known that $m_7(\eta)\not=0$ from (3.1) and $u_5u_6-exp(ku)\not=0$ from the initial conditions (2.8). Therefore, the following system of equations is equivalent to (3.8)
$$\begin{array}{l}
(m_7u_5u_6(u_4u_6-u_1u_5)+m_7u_4u_5(u_5^2-u_4^2)+m_7exp(ku)(u_1u_5-u_4u_6)+m_6u_7(u_4^2-u_5^2)=0\\[4mm]
u_1^2u_5-2u_1u_4u_6+u_4^2u_5-u_5^3+u_5u_6^2=0\\[4mm]
-u_1u_4^2u_6+u_1u_5^2u_6+u_4^3u_5-u_4u_5^3-u_1u_5exp(ku)+u_4u_6exp(ku)=0\\[4mm]
-u_1^2u_4+2u_1u_5u_6+u_4^3-u_4u_5^2-u_4u_6^2=0
\end{array}
\eqno{(3.9)}$$

{\small\it Remark 1. When we perform the operation of eliminating indeterminates in a system of equations,according to the theorem of zero structure of algebraic varieties [16], the coefficient of the term with the highest degree of the indeterminate in the main polynomial equation(to be eliminated in the other polynomials)should not be identified with zero. In the event it is identified with zero, we should add the coefficient into the equations to produce a new system of equations.Otherwise, it is possible to lose some solutions. In our cases,we can discover that the coefficients of the terms which are eliminated are not identified with zero thanks to the initial conditions and the nondegenerate conditions we have set.}

From the argument above, we have known the system of equation (3.3) is equivalent to two groups of equations. The first is (3.4) and (3.9) called Baxter case. The second is (3.4) and (3.6) called free-fermion case. We will discuss them respectly.

{\it 3-1.Baxter-type solutions}

We consider the first case in $k\not=0$ case now i.e. (3.4) and (3.9). If we differentiate the second  and the fourth equations in (3.9) and take $u=0,\xi=\eta$ and substitute the initial conditions into the results, we can prove
$$\begin{array}{ll}
m_5(\eta)=m_6(\eta)&m_1(\eta)=m_4(\eta)
\end{array}
\eqno{(3-1.1)}$$
By eliminating $u_4$ we can factorize the last equation in (3.9) to be
$$u_5^2(u_6-u_5)(u_6+u_5)(u_6-u_1)(u_6+u_1)exp(ku)=0
\eqno{(3-1.2)}$$
So
$$a_5(u,\xi,\eta)=a_6(u,\xi,\eta)
\eqno{(3-1.3a)}$$
or
$$u_5(u,\xi,\eta)=-u_6(u,\xi,\eta)
\eqno{(3-1.3b)}$$
will hold because of the initial conditions (2.8).

If $u_5=-u_6$, together with the second equation of (3.9) we have $u_1=-u_4$ which is impossible for the initial conditions (2.8). Then, we have only $u_5=u_6$. Substituting (3-1.3a) into the second equation of (3.9) we can obtain
$$a_1(u,\xi,\eta)=a_4(u,\xi,\eta)
\eqno{(3-1.4)}$$
Combining (3-1.3a),(3-1.4) and the first equation of (3.9), we can get
$$(m_7(\eta)u_1u_5-m_5(\eta)u_7)(u_5^2-u_1^2)=0
\eqno{(3-1.5)}$$
Then the following equation is correct for the initial conditions(2.8)
$$m_7(\eta)u_1u_5=m_5(\eta)u_7
\eqno{(3-1.6)}$$
So, we have
$$m_5(\eta)\not=0
\eqno{(3-1.7)}$$
because of (3.1).
From (2.11a), (3-1.3a) and (3-1.4) we get
$$\frac{\partial a_7(u,\xi,\eta)}{\partial u}=m_7(\eta)(a_1^2(u,\xi,\eta)-a_5^2(u,\xi,\eta))$$
Substituting (3.2) into it 
$$F_7(\xi,\eta)exp(ku)=F_7(\eta,\eta)(u_1^2-u_5^2)
\eqno{(3-1.8)}$$
then differentiating the result above with respect to u and letting $u=0,\xi=\eta$ it is easy to obtain
$$m_1(\eta)=\frac{k}{2}
\eqno{(3-1.9)}$$
From (3.4), (3-1.1), (3-1.3a), (3-1.4) and (3-1.9) we also have
$$m_5(\eta)(u_5^2+u_1^2-exp(ku))=0$$
So there must be
$$u_5^2+u_1^2=exp(ku)
\eqno{(3-1.10)}$$
by (3-1.7). But combining (3-1.8) and (3-1.10) by letting $\xi=\eta$ we have
$$a_1^2(u,\eta,\eta)-a_5^2(u,\eta,\eta)=a_1^2(u,\eta,\eta)+a_5^2(u,\eta,\eta)=exp(ku)$$
It is to show that

$a_5(u,\eta,\eta)=0$ i.e. $m_5(\eta)=0$\\
which has discrepancy with (3-1.7). So, there is no solution in this case.

{\it 3-2. Free-fermion-type solutions}

We consider the second case now i.e. (3.4) and (3.6). If we differentiating (3.6) with respect to u and then let $u=0,\eta=\xi$ we can get
$$m_1(\eta)+m_4(\eta)=k
\eqno{(3-2.1)}$$
Substituting (3.6) into (3.4) we obtain
$$m_7(\eta)(u_1^2+u_6^2-u_4^2-u_5^2)=2(m_1(\eta)-m_4(\eta))u_7\eqno{(3-2.2a)}$$
$$m_7(\eta)(u_1u_6+u_4u_5)=(m_6(\eta)+m_5(\eta))u_7
\eqno{(3-2.2b)}$$
If we differentiate the system of equations with respect to $u$ after eliminating $w_i(i=1,4,\ldots,7)$ from (2.7) and then set $u=0,\eta=\xi$ and replace $v$   
as well as $\lambda$ by $u$ as well as $\eta$ we can get
$$\begin{array}{l}
m_6(\xi)(u_1u_4+u_5u_6-exp(ku))+(k-m_1(\xi)-m_4(\xi))u_1u_6=0\\[4mm]
m_5(\xi)(u_1u_4+u_5u_6-exp(ku))+(k-m_1(\xi)-m_4(\xi))u_4u_6=0\\[4mm]
m_7(\xi)(u_1^2-u_4^2)-(m_5(\xi)+m_6(\xi))(u_1u_6-u_4u_5)u_7+2(m_4(\xi)-m_1(\xi))u_1u_4u_7=0\\[4mm]
m_7(\xi)(u_1u_6+u_4u_5)-(m_5(\xi)+m_6(\xi))(u_1^2+u_5^2)u_7-2(m_4(\xi)-m_1(\xi))u_1u_5u_7=0\\[4mm]
m_7(\xi)(u_1u_5+u_4u_6)-(m_5(\xi)+m_6(\xi))u_7exp(ku)=0\\[4mm]
m_7(\xi)u_1(u_6^2-u_1^2)-(m_5(\xi)+m_6(\xi))u_5u_7exp(ku)-2(m_4(\xi)-m_1(\xi))u_1u_7exp(ku)+m_7(\xi)u_4exp(ku)=0\\[4mm]
m_7(\xi)u_1(u_5^2-u_4^2)-(m_5(\xi)+m_6(\xi))u_5u_7exp(ku)+m_7(\xi)u_4exp(ku)=0\\
\end{array}
\eqno{(3-2.3)}$$
From the equations above, we can obtain
$$m_7(\xi)(u_1^2+u_5^2-u_4^2-u_6^2)=2(m_1(\xi)-m_4(\xi))u_7exp(ku)
\eqno{(3-2.4a)}$$
$$m_7(\xi)(u_1u_5+u_4u_6)=(m_5(\xi)+m_6(\xi))u_7exp(ku)
\eqno{(3-2.4b)}$$

If we set $\eta=\xi$ in (3-2.2) and (3-2.4), we can see from these two systems of equations that the following equation is sound
$$(\hat u_1\hat u_5+\hat u_4\hat u_6)(\hat u_1^2+\hat u_6^2-\hat u_4^2-\hat u_5^2)=(\hat u_1\hat u_6+\hat u_4\hat u_5)(\hat u_1^2+\hat u_5^2-\hat u_4^2-\hat u_6^2)$$
We can write it in the form of factorization
$$(\hat u_5-\hat u_6)(\hat u_1+\hat u_4)(\hat u_1-\hat u_4+\hat u_5+\hat u_6)(-\hat u_1+\hat u_4+\hat u_5+\hat u_6)=0$$
Here we have used the denotation
$$\hat u_i=a_i(u,\xi,\xi)$$
Then thanks to the initial conditions (2.8) and the solution transformation {\bf A}, we get
$$a_5(u,\xi,\xi)=a_6(u,\xi,\xi)
\eqno{(3-2.5a)}$$
or
$$a_1(u,\xi,\xi)=a_4(u,\xi,\xi)+a_5(u,\xi,\xi)+a_6(u,\xi,\xi)
\eqno{(3-2.5b)}$$
without losing generality. We will discuss them seperately.

{\it (A). $a_1(u,\xi,\xi)=a_4(u,\xi,\xi)+a_5(u,\xi,\xi)+a_6(u,\xi,\xi)$ case}

Equation (3-2.5b) means
$$m_1-m_4=m_5+m_6$$
Substituting it into (3-2.2) and (3-2.3) we have
$$\begin{array}{l}
(u_1-u_5)^2=(u_4+u_6)^2\\[4mm]
(u_1-u_6)^2=(u_4+u_5)^2
\end{array}$$
Combining the initial conditions (2.8), it is to say
$$u_1=u_4+u_5+u_6
\eqno{(3-2-1.1)}$$
With the free-fermion condition (3.6), we can affirm
$$\begin{array}{ll}
\check m_1(\eta)+\check m_4(\eta)=0&\hat m_1(\xi)+\hat m_4(\xi)=0\\[4mm]
\check m_1(\eta)=\check m_4(\eta)+\check m_5(\eta)+\check m_6(\eta)&\hat m_1(\xi)=\hat m_4(\xi)+\hat m_5(\xi)+\hat m_6(\xi)
\end{array}
\eqno{(3-2-1.2)}$$
Substituting them into the fifth equation of (2.10c) as well as its counterpart and the fourth equation of (2.10d) as well as its counterpart, we will get
$$\begin{array}{l}
\frac{\partial u_7}{\partial \eta}=\check m_7(\eta) \exp(ku)\\[4mm]
\frac{\partial u_7}{\partial \xi}=\hat m_7(\xi)
\end{array}
\eqno{(3-2-1.3)}$$
Then, from the solution form of $u_7$ (3.2) we obtain
$$a_7(u,\xi,\eta)=\exp(H(\eta))(\exp(ku)-\exp(H(\xi)-H(\eta))
\eqno{(3-2-1.4)}$$
The following work is to substituting (3-2-1.1) into the sixth equation of (2.10a) as well as its counterpart and (3-2.2b)
$$(u_1^2-u_5^2)m_7(\eta)-(2m_1(\eta)-k)u_7-u_7^{'}=0
\eqno{(3-2-1.5a)}$$
$$(u_4^2-u_6^2)m_7(\eta)+(2m_1(\eta)-k)u_7-u_7^{'}=0
\eqno{(3-2-1.5b)}$$
$$u_1 u_4+u_5 u_6=\exp(ku)\eqno{(3-2-1.5c)}$$
$$(u_1 u_6+u_4 u_5)m_7(\eta)=(2m_1(\eta)-k)u_7\eqno{(3-2-1.5d)}$$
Substituting (3-2-1.4) into the sixth equation of (2.10b) as well as its counterpart and (3-2.4b) there will be
$$(u_1^2-u_6^2)m_7(\xi)-(2m_1(\xi)-k)u_7\exp(ku)-u_7^{'}\exp(ku)=0
\eqno{(3-2-1.6a)}$$
$$(u_4^2-u_5^2)m_7(\xi)+2m_1(\xi)u_7\exp(ku)-u_7^{'}\exp(ku)=0
\eqno{(3-2-1.6b)}$$
$$u_1 u_4+u_5 u_6=\exp(ku)\eqno{(3-2-1.6c)}$$
$$(u_1 u_5+u_4 u_6)m_7(\xi)=(2m_1(\xi)-k)u_7\exp(ku)\eqno{(3-2-1.6d)}$$
First, we assum 
$$
k\not=2m_1(\xi)$$
Then by these two systems of equations, we easily obtain
$$\displaystyle\frac{u_4}{u_1}=\displaystyle\frac{2k(m_1(\xi)+m_1(\eta))\exp(H(\xi))-2k^2\exp(H(\xi))-2u_7(k-2m_1(\xi))(k-2m_1(\eta))}{2k(m_1(\xi)+m_1(\eta))\exp(H(\xi))-2k^2\exp(H(\xi))-2u_7m_1(\eta)(k-2m_1(\xi))}
\eqno{(3-2-1.7a)}$$
$$\displaystyle\frac{u_5}{u_6}=\displaystyle\frac{2k(m_1(\xi)-m_1(\eta))\exp(H(\xi))-2m_1(\eta)u_7(k-2m_1(\xi))}{2k(m_1(\eta)-m_1(\xi))\exp(H(\xi))+2u_7(k-m_1(\eta))(k-2m_1(\xi))}
\eqno{(3-2-1.7b)}$$
We denote here
$$\begin{array}{l}
H_1=2k(m_1(\xi)+m_1(\eta))\exp(H(\xi))-2k^2\exp(H(\xi))-2u_7m_1(\eta)(k-2m_1(\xi))\\[4mm]
H_4=2k(m_1(\xi)+m_1(\eta))\exp(H(\xi))-2k^2\exp(H(\xi))-2u_7(k-2m_1(\xi))(k-2m_1(\eta))\\[4mm]
H_5=2k(m_1(\xi)-m_1(\eta))\exp(H(\xi))-2m_1(\eta)u_7(k-2m_1(\xi))\\[4mm]
H_6=2k(m_1(\eta)-m_1(\xi))\exp(H(\xi))+2u_7(k-m_1(\eta))(k-2m_1(\xi))
\end{array}$$
Then we can assum
$$\begin{array}{ll}
u_1=H_1X&u_4=H_4X\\[4mm]
u_5=H_5Y&u_6=H_6Y
\end{array}
\eqno{(3-2-1.8)}$$
Furthermore, from (3-2-1.5), we can see
$$\begin{array}{l}
X^2(H_1^2-H_4^2)=u_1^2-u_4^2=\displaystyle\frac{u_7^{'}+(2m_1(\eta)-k)u_7}{m_7(\eta)}-\displaystyle\frac{u_7^{'}-2m_1(\xi)u_7}{m_7(\xi)}\exp(ku)\\[4mm]
Y^2(H_6^2-H_5^2)=u_6^2-u_5^2=\displaystyle\frac{u_7^{'}-(k-2m_1(\eta)u_7}{m_7(\eta)}-\displaystyle\frac{u_7^{'}-2(k-m_1(\xi))u_7}{m_7(\xi)}\exp(ku)
\end{array}
\eqno{(3-2-1.9)}$$
where $m_1(\xi), H(\xi)$ are arbitary functions of coloured parameter, $k$ is a non-zero complex constant and $m_7(\xi)=k\exp(H(\xi))$ defined by (3-2-1.4). Now, we can calcuate out $X$ and $Y$ from the equations above, which means we can get $u_1, u_4, u_5$ and $u_6$ by (3-2-1.8).

If we setting $\eta=\xi$ in this solution, we immediately get a corresponding solution for the seven-vertex pure spectral YBE 
$$\begin{array}{l}
a_1(u)=\alpha (\exp(ku)-1)+1\\[4mm]
a_4(u)=1-(\alpha -1)(\exp(ku)-1)\\[4mm]
a_5(u)=\alpha (\exp(ku)-1)\\[4mm]
a_6(u)=(\alpha-1)(\exp(ku)-1)\\[4mm]
a_7(u)=\beta (\exp(ku)-1)
\end{array}
\eqno{(3-2-1.10)}$$
which has no correspondence in [25] and where $\alpha, \beta$ and $k$ are complex constants. 

Now, let's consider 
$$k=2m_1(\xi)$$
case where we can see $m_1(\xi)$ is a complex constant.In this case, from (3-2.2) and (3-2.4) we can get
$$\begin{array}{l}
u_1^2+u_6^2-u_4^2-u_5^2=0\\[4mm]
u_1u_6+u_4u_5=0\\[4mm]
u_1^2+u_5^2-u_4^2-u_6^2=0\\[4mm]
u_1u_5+u_4u_6=0
\end{array}$$
which is eqivalent to the following equations
$$\begin{array}{ll}
u_6=-u_5&u_1=u_4
\end{array}
\eqno{(3-2-1.11)}$$
for the initial condition (2.8).

Substituting the equation above into (2.10a) and (2.10b) we can get
$$\begin{array}{l}
\frac{\partial}{\partial u}a_1(u,\xi,\eta)=\frac{k}{2}a_1(u,\xi,\eta)+m_5(\eta)u_5\\[4mm]
\frac{\partial}{\partial u}a_5(u,\xi,\eta)=\frac{k}{2}a_5(u,\xi,\eta)+m_5(\eta)u_1\\[4mm]
\end{array}
\eqno{(3-2-1.12a)}$$
$$\begin{array}{l}
\frac{\partial}{\partial u}a_1(u,\xi,\eta)=\frac{k}{2}a_1(u,\xi,\eta)+m_5(\xi)u_5\\[4mm]
\frac{\partial}{\partial u}a_5(u,\xi,\eta)=\frac{k}{2}a_5(u,\xi,\eta)+m_5(\xi)u_1\\[4mm]
\end{array}
\eqno{(3-2-1.12b)}$$
Then, we can see from them that $m_5(\xi)=\beta$ is a complex constant. So, we can rewrite (3-2-1.12) as
$$\begin{array}{l}
\frac{\partial}{\partial u}a_1(u,\xi,\eta)=\frac{k}{2}a_1(u,\xi,\eta)+\beta u_5\\[4mm]
\frac{\partial}{\partial u}a_5(u,\xi,\eta)=\frac{k}{2}a_5(u,\xi,\eta)+\beta u_1\\[4mm]
\end{array}
\eqno{(3-2-1.13)}$$
On the other hand, from (3.6) and (3-2-1.11) we can obtain
$$\check m_1(\eta)=\check m_4(\eta)=0
\eqno{(3-2-1.14)}$$
Then substituting (3-2-1.11) and (3-2-1.14) into (2.10c) we have
$$\begin{array}{l}
\frac{\partial}{\partial \eta}a_1(u,\xi,\eta)=\check m_5(\eta)a_5(u,\xi,\eta)\\[4mm]
\frac{\partial}{\partial \eta}a_5(u,\xi,\eta)=\check m_5(\eta)a_1(u,\xi,\eta)\\
\end{array}
\eqno{(3-2-1.15)}$$
It is easy to get the solution from (3-2-1.13) and (3-2-1.15) by using the initial condition (2.8)

$$\begin{array}{l}
a_1(u,\xi,\eta)=a_4(u,\xi,\eta)=exp(ku/2)\cosh(\beta u+F(\xi)-F(\eta))\\[4mm]
a_2(u,\xi,\eta)=1\\[4mm]
a_3(u,\xi,\eta)=exp(ku)\\[4mm]
a_5(u,\xi,\eta)=-a_6(u,\xi,\eta)=exp(ku/2)\sinh(\beta u+F(\xi)-F(\eta))exp\\[4mm]
a_7(u,\xi,\eta)=F_7(\eta)exp(ku)-F_7(\xi)
\end{array}
\eqno{(3-2-1.16)}$$
where $k$ is nonzero complex constant, $\beta$ is a arbitary complex constant and $F(\xi),F_7(\xi)$ are two arbitary function of coloured parameter. 

Taking the coloured parameters
$\xi=\eta$ in this solution will degenerate into one of the solutions of pure spectral YBE which is the same as the third case given by [25].

{\it (B). $a_5(u,\xi,\xi)=a_6(u,\xi,\xi)$ case}

Equation (3-2.5a) means
$$m_5(\xi)=m_6(\xi)\eqno{(3-2-2.1)}$$
Setting $\eta=\xi$ substituting (3-2.5a) and (3-2-2.1) into (3-2.2b) and (3-2.4b) we have
$$(m_5(\xi)+m_6(\xi))\hat u_7=(m_5(\xi)+m_6(\xi))\hat u_7exp(ku)$$
Then we obtain
$$m_5(\xi)=m_6(\xi)=0
\eqno{(3-2-2.2)}$$
So, from (3-2.2b) and (3-2.4b) it is easy to show 
$$m_7(\eta)(u_1u_6+u_4u_5)=m_7(\xi)(u_1u_5+u_4u_6)=0$$
i.e.\\
$(u_1-u_4)(u_5-u_6)=0$ and $u_1u_6+u_4u_5=0$\\
If we take $u_5=u_6$ there must be $u_1+u_4=0$ which disagrees with the initial conditions (2.8) in the above equations. Therefore, we can only choose
$$\begin{array}{ll}
u_1=u_4&u_5=-u_6
\end{array}
\eqno{(3-2-2.3)}$$
which has been discussed by us. So, there is no new solution.

\section{The solutions of equation (2.1) in the case of $k=0$}

Now equation (2.11) changs to:
$$2u_7^{'}=m_7(\eta)(u_1^2+u_4^2-u_5^2-u_6^2)=m_7(\xi)(u_1^2+u_4^2-u_5^2-u_6^2)
\eqno{(4.1)}$$
It is obviously that
$$m_7(\xi)=m_7(\eta)=\alpha
\eqno{(4.2)}$$
thanks to the initial conditions (2.8). Here $\alpha$ is a complex constant independent of coloured parameters.

{\it proposition 4.1} In the case of $k=0$, there is at least one between $m_5(\xi)$(or $m_6(\xi)$) and $m_7(\xi)$ which is not zero identically. Otherwise, the solution will be independent of spectral parameter.

{\it Proof.} If $m_7(\xi)=0$ we have $u_7^{'}=0$ because of (4.1). Now substituting them into the fourth and fifth equation of (2.10a)($k=0$),
we have
$$\begin{array}{l}
u_1^{'}=u_6m_6(\eta)-u_1m_1(\eta)\\[4mm]
u_6^{'}=u_6m_1(\eta)-u_1m_5(\eta)
\end{array}
\eqno{(4.3)}$$
which means
$$\begin{array}{ll}
m_1(\eta)=0&m_5(\eta)+m_6(\eta)=0
\end{array}
\eqno{(4.4)}$$
by differentiating both sides of (4.3) and letting $u=0,\eta=\xi$ Substituting the results into the first three equations of (2.10a)
we obtain
$$\begin{array}{ll}
u_1^{'}=-m_5(\eta)u_6&u_4^{'}=m_5(\eta)u_5\\
u_5^{'}=-m_5(\eta)u_4&u_6^{'}=-m_5(\eta)u_1
\end{array}$$
So
$$\begin{array}{ll}
u_i^{''}=m_5^2(\eta)u_i & (i=1,4,5,6)
\end{array}$$
will holds.
If we start from (2.10b) to do the same procedure as the above, we can get
$$\begin{array}{ll}
u_i^{''}=m_5^2(\xi)u_i& (i=1,4,5,6)
\end{array}$$ 
Then we can know
$$m_5(\eta)=m_5(\xi)=\beta$$
where $\beta$ is a complex constant independent of coloured parameter.
Now we come to the conclusion of proposition 4.1.

{\it Remark 2. From (2.10) we can observe that if $k=0$:
\begin{itemize}
\item (2.10a) is the same as (2.10c) except replace $m_i(\eta)$ by $ \check m_i(\eta)$.
\item (2.10b) is the same as (2.10d) except replace $m_i(\eta)$ by $ \hat m_i(\eta)$.
\item (2.10a) is the same as (2.10b) if we interchang sub-indices 5 and 6
and replace $m_i(\eta)$ by $m_i(\xi)$.
\item (2.10c) is the same as (2.10d) if we interchang sub-indices 5 and 6
and replace $\check m_i(\eta)$ by $\hat m_i(\xi)$.
\end{itemize}}
In fact, those tricks are always sound in the following of this paper and we will ofen employ them. We call  the kind of transformation from (2.10a) to (2.10b) as symmetric operation.

As the same as $k\not=0$ case, we have:
$$\begin{array}{l}
m_5+(m_1+m_4)u_1u_5-m_5(u_1u_4+u_5u_6)=0\\[4mm]
m_7u_1u_5(u_4^2-u_1^2+u_5^2-u_6^2)+2m_5u_7(u_1u_4+u_5u_6-1))-4m_4u_1u_5u_7=0\\[4mm]
m_7u_1(u_4^2u_5-u_5^2u_6+u_6-u_7(m_5+m_6u_1u_4-m_5u_5u_6)=0
\end{array}
\eqno{(4.5)}$$
which contains $m_1,m_4$ and $m_5$. And
$$u_1u_4+u_5u_6=1
\eqno{(4.6)}$$
which is the free-fermion condition [12] or
$$\begin{array}{l}
m_7u_5u_6(u_4u_6-u_1u_5)+m_7u_4u_5(u_5^2-u_4^2)+m_7(u_1u_5-u_4u_6)+m_6u_7(u_4^2-u_5^2)=0\\
u_1^2u_5-2u_1u_4u_6+u_4^2u_5-u_5^3+u_5u_6^2=0\\
-u_1u_4^2u_6+u_1u_5^2u_6+u_4^3u_5-u_4u_5^3-u_1u_5+u_4u_6=0\\
-u_1^2u_4+2u_1u_5u_6+u_4^3-u_4u_5^2-u_4u_6^2=0
\end{array}
\eqno{(4.7)}$$
Here we have used proposition 4.1.

Now there are also two cases: the first is (4.5) as well as (4.7), Baxter case and the second is (4.5) as well as (4.6), the free-fermion case.

{\it 4-1. Baxter-type solutions}

From (4.7), it is easy to obtain
$$\begin{array}{lll}
m_5(\eta)=m_6(\eta)&m_1(\eta)=m_4(\eta)&m_7=\alpha\not=0
\end{array}
\eqno{(4-1.1)}$$
From (4.7), by eliminating $u_4$, we also have
$$u_5^2(u_6-u_5)(u_6+u_5)(u_6-u_1)(u_6+u_1)=0
\eqno{(4-1.2)}$$
So
$$a_5(u,\xi,\eta)=a_6(u,\xi,\eta)
\eqno{(4-1.3a)}$$
or
$$u_5(u,\xi,\eta)=-u_6(u,\xi,\eta)
\eqno{(4-1.3b)}$$
will hold because of the initial conditions (2.8).

If $u_5=-u_6$, together with the second equation of (4.7) we have $u_1=-u_4$ which is impossible for the initial conditions (2.8). Then, we have only $u_5=u_6$. Substituting (4-1.3a) into the second equation of (4.7) we can obtain
$$a_1(u,\xi,\eta)=a_4(u,\xi,\eta)
\eqno{(4-1.4)}$$
Combining (4-1.3a), (4-1.4) and the first equation of (4.7), we can get
$$(m_7(\eta)u_1u_5-m_5(\eta)u_7)(u_5^2-u_1^2)=0
\eqno{(4-1.5)}$$
Then the following equation is correct for the initial conditions (2.8) and (4.2)
$$\alpha u_1u_5=m_5(\eta)u_7
\eqno{(4-1.6)}$$
Using the symmetric operation, we also get
$$\alpha u_1u_6=\alpha u_1u_5=m_6(\xi)u_7=m_5(\xi)u_7$$
So, there has
$$m_5(\xi)=\beta
\eqno{(4-1.7)}$$
where $\beta$ is coplex constant independent of coloured parameter.
There also is
$$u_7=\frac{\alpha}{\beta}u_1u_5
\eqno{(4-1.8)}$$

Substituting (4-1.3a) and (4-1.4) into (2.10a) and (4.5), we can obtain the following conclusion through some calculation:
$$\begin{array}{l}
(u_5^{'})^2=\beta ^2-(\beta ^2-m_1(\eta)^2)u_5^2\\[4mm]
(u_1^{'})^2=\beta ^2-(\beta ^2-m_1(\eta)^2)u_1^2\\[4mm]
\beta ^2(1-u_5^2-u_1^2)+2\beta m_1(\eta)u_1u_5=0
\end{array}
\eqno{(4-1.9)}$$
If we use the symmetric operation, we also get
$$\begin{array}{l}
(u_5^{'})^2=\beta ^2-(\beta ^2-m_1(\xi)^2)u_5^2\\[4mm]
(u_1^{'})^2=\beta ^2-(\beta ^2-m_1(\xi)^2)u_1^2\\[4mm]
\beta ^2(1-u_5^2-u_1^2)+2\beta m_1(\xi)u_1u_5=0
\end{array}
\eqno{(4-1.10)}$$
So we can affirm
$$m_1(\xi)=\gamma
\eqno{(4-1.11)}$$
Then, (4-1.9) can be rewritten
$$\begin{array}{l}
(u_5^{'})^2=\beta ^2-(\beta ^2-\gamma ^2)u_5^2\\[4mm]
(u_1^{'})^2=\beta ^2-(\beta ^2-\gamma ^2)u_1^2\\[4mm]
\beta ^2(1-u_5^2-u_1^2)+2\beta \gamma u_1u_5=0
\end{array}
\eqno{(4-1.12)}$$
Substituting (4-1.3a) and (4-1.4) into (2.10c) and (4.5), we can obtain the following conclusion through some calculation:
$$\begin{array}{l}
(\frac{\partial \check u_5}{\partial \eta})^2=\check m_5(\eta) ^2-(\check m_5(\eta) ^2-\check m_1(\eta)^2)\check u_5^2\\[4mm]
(\frac{\partial \check u_1}{\partial \eta})^2=\check m_5(\eta) ^2-(\check m_5(\eta) ^2-\check m_1(\eta)^2)\check u_1^2
\end{array}
\eqno{(4-1.13)}$$
and from the last equation of (4-1.9) we also have
$$\frac{(\check m_5(\eta))^2-(\check m_1(\eta))^2}{(\check m_5(\eta))^2}=1-\frac{(u_5^2+u_1^2-1)^2}{4u_1^2u_5^2}=\frac{\beta ^2-\gamma ^2}{\beta ^2}
\eqno{(4-1.14)}$$
So, (4-1.13) can be rewritten
$$\begin{array}{l}
(\frac{\partial \check u_5}{\partial \eta})^2=\check m_5(\eta) ^2(1-\frac{\beta ^2-\gamma ^2}{\beta ^2}\check u_5^2)\\[4mm]
(\frac{\partial \check u_1}{\partial \eta})^2=\check m_5(\eta) ^2(1-\frac{\beta ^2-\gamma ^2}{\beta ^2}\check u_1^2)
\end{array}
\eqno{(4-1.15)}$$
Combining (4-1.9), (4-1.15), (4-1.8) and the initial conditions (2.8), we can immediately write down the  solutions of (2.1) in this case\\

{\it subcase of $\beta ^2=\gamma ^2$}(we can let $\beta =\gamma$ according to the solution transformation {\bf C}).
$$\begin{array}{l}
a_1(u,\xi,\eta)=a_4(u,\xi,\eta)=\beta u+F(\xi)-F(\eta)+1\\[4mm]
a_2(u,\xi,\eta)=a_3(u,\xi,\eta)=1\\[4mm]
a_5(u,\xi,\eta)=a_6(u,\xi,\eta)=\beta u+F(\xi)-F(\eta)\\[4mm]
a_7(u,\xi,\eta)=\frac{\alpha}{\beta}(\beta u+F(\xi)-F(\eta)+1)(\beta u+F(\xi)-F(\eta))
\end{array}
\eqno{(4-1.16a)}$$
{\it subcase of $\beta ^2\not=\gamma ^2$}.
$$\begin{array}{l}
a_1(u,\xi,\eta)=a_4(u,\xi,\eta)=\frac{cos(\sqrt{\beta ^2-\gamma ^2}u+F(\xi)-F(\eta)-\theta)}{cos\theta}\\
a_2(u,\xi,\eta)=a_3(u,\xi,\eta)=1\\
a_5(u,\xi,\eta)=a_6(u,\xi,\eta)=\frac{sin(\sqrt{\beta ^2-\gamma ^2}u+F(\xi)-F(\eta))}{cos\theta}\\
a_7(u,\xi,\eta)=\frac{\alpha}{\beta cos^2\theta} cos(\sqrt{\beta ^2-\gamma ^2}u+F(\xi)-F(\eta)-\theta) sin(\sqrt{\beta ^2-\gamma ^2}u+F(\xi)-F(\eta))
\end{array}
\eqno{(4-1.16b)}$$
where the defination of $\theta$ is
$$\begin{array}{ll}
sin\theta= \frac{\gamma}{\beta}&cos\theta =\frac{\sqrt{\beta ^2-\gamma ^2}}{\beta}
\end{array}$$
By letting $\beta \rightarrow \gamma$, we can see that (4-1.16a) is the degenerate
form of 4-1.16b.

Furthermore, if we let $\xi=\eta$ in (4-1.16b) which means one of the  seven-vertex-type solutions of YBE without coloured parameters, we can immediately find it is the same as the first type solution which was obtained in [25].

{\it 4-2. Free-fermion-type solutions}

Now, let's consider the second case of $k=0$ i.e. equation (4.5) and (4.6). From (4.6) we can affirm 
$$m_1(\eta)+m_4(\eta)=0
\eqno{(4-2.1)}$$
by differentiating it and then setting $u=0,\eta=\xi$.

Substituting (4.6) into (4.5) we get
$$u_1u_4+u_5u_6=1\eqno{(4-2.2a)}$$
$$\alpha (u_1^2+u_6^2-u_4^2-u_5^2)=2(m_1(\eta)-m_4(\eta))u_7\eqno{(4-2.2b)}$$
$$\alpha (u_1u_6+u_4u_5)=(m_6(\eta)+m_5(\eta))u_7
\eqno{(4-2.2c)}$$
Using the symmetric operation, there are
$$u_1u_4+u_5u_6=1\eqno{(4-2.3a)}$$
$$\alpha (u_1^2+u_5^2-u_4^2-u_6^2)=2(m_1(\xi)-m_4(\xi))u_7\eqno{(4-2.3b)}$$
$$m_7\alpha (u_1u_5+u_4u_6)=(m_6(\xi)+m_5(\xi))u_7
\eqno{(4-2.3c)}$$
If we set $\eta=\xi$ in (4-2.2b) and (4-2.3b), we can get
$$a_6(u,\xi,\xi)^2=a_5(u,\xi,\xi)^2\eqno{(4-2.4a)}$$
i.e.
$$m_5(\eta)^2=m_6(\eta)^2\eqno{(4-2.4b)}$$
From (4-2.2) and the last equation in (2.10a) as well as its counterpart, we can write down
$$(u_7^{'})^2=\alpha ^2-((m_5(\eta)+m_6(\eta))^2-4m_1(\eta)^2)u_7^2\eqno{(4-2.5a)}$$
Using the symmetric operation
$$(u_7^{'})^2=\alpha ^2-((m_5(\xi)+m_6(\xi))^2-4m_1(\xi)^2)u_7^2\eqno{(4-2.5b)}$$
So, we can obtain
$$\delta ^2=(m_5(\xi)+m_6(\xi))^2-4m_1(\xi)^2\eqno{(4-2.6)}$$
where $\delta$ is a complex constant in dependent of coloured parameters. In the following, we will respectly discuss the two cases $m_5(\xi)=m_6(\xi)$ and $m_5(\xi)=-m_6(\xi)$

{\it Subcase 4-2-1. $m_5(\xi)=-m_6(\xi)$ case}
From (4-2.2) and (4-2.3) we have
$$\begin{array}{ll}
u_1=u_4&u_5=-u_6\\
u_1^2-u_5^2=1
\end{array}
\eqno{(4-2-1.1)}$$
Together with (4-2.1), there will be
$$m_1(\xi)=m_4(\xi)=0\eqno{(4-2-1.2)}$$
So, from (2.10a)
$$\begin{array}{l}
u_7^{'}=\alpha\\[4mm]
(u_5^{'})^2=(m_5(\eta))^2(1+u_5^2)\\[4mm]
(u_1^{'})^2=(m_5(\eta))^2(u_1^2-1)
\end{array}
\eqno{(4-2-1.3)}$$
will be hold. Using the symmetric operation, we also can affirm
$$m_5(\eta)=\beta\eqno{(4-2-1.4)}$$
So, (4-2-1.3) can be rewritten as
$$\begin{array}{l}
u_7^{'}=\alpha\\
(u_5^{'})^2=\beta ^2(1+u_5^2)\\
(u_1^{'})^2=\beta ^2(u_1^2-1)
\end{array}
\eqno{(4-2-1.5a)}$$

Thanks to the remark 2. if we work the same process to equaton (2.10c), we can obtain
$$\begin{array}{l}
\frac{\partial u_7}{\partial \eta}=(\check m_7(\eta))^2\\[4mm]
(\frac{\partial u_1}{\partial \eta})^2=(\check m_5(\eta))^2(u_1^2-1)\\[4mm]
(\frac{\partial u_5}{\partial \eta})^2=(\check m_5(\eta))^2(u_5^2+1)
\end{array}
\eqno{(4-2-1.5b)}$$
Now, from (4-1-1.5) and the initial conditions (2.8) the solution of (2.1) in this case is
$$\begin{array}{l}
a_1(u,\xi,\eta)=a_4(u,\xi,\eta)=cosh(\beta u+F(\xi)-F(\eta))\\[4mm]
a_2(u,\xi,\eta)=a_3(u,\xi,\eta)=1\\[4mm]
a_5(u,\xi,\eta)=-a_6(u,\xi,\eta)=sinh(\beta u+F(\xi)-F(\eta))\\[4mm]
a_7(u,\xi,\eta)=\alpha u+F(\xi)-F(\eta)
\end{array}
\eqno{(4-2-1.6)}$$

Setting $\eta=\xi$, we can get a solution for the pure YBE which has been lost in [25].

{\it Subcase 4-2-2. $m_5(\xi)=m_6(\xi)$ case}

Now, equation (4-2.5) and (4-2.6) chang to be
$$(u_7^{'})^2=\alpha ^2-\delta ^2u_7^2\eqno{(4-2-2.1)}$$
and
$$\delta ^2=4m_5(\xi)^2-4m_1(\xi)^2\eqno{(4-2-2.2)}$$
And from (4-2.2c) we can know in this case
$$\alpha\not=0\eqno{(4-2-2.3)}$$
otherwise, there will be $m_5(\xi)=0$ which has discrepancy with proposition 4.1. If we try to solve equation (4-2-2.1), we can find there have two branches which is related to $\delta=0$ and $\delta\not=0$.

{\it (A). $\delta=0$ subcase}

We can see that $\delta=0$ means $m_1(\xi)^2=m_5(\xi)^2$. Thanks to the solution transformation {\bf C} and equation (4-2.1), we can set
$$m_1(\xi)=m_5(\xi)=m_6(\xi)=-m_4(\xi)\eqno{(4-2-2a.1)}$$
At this time, (4-2-2.1) changes to be
$$u_7^{'}=\alpha \eqno{(4-2-2a.2)}$$
thanks to the solution transformation {\bf C}.
 Now, let's consider (4-2.2b) and (4-2.2c). Using (4-2-2a.1) we can get
$$u_1^2+u_6^2-u_4^2-u_5^2-2u_1u_6-2u_4u_5=0$$
Considering (4-2.3b) and (4-2.3c), using (4-2-2a.1) we can also get
$$u_1^2+u_5^2-u_4^2-u_6^2-2u_1u_5-2u_4u_6=0$$
From these two equations, it is easy to obtain
$$2(u_1+u_4)(u_1-u_4-u_5-u_6)=0$$
which means
$$u_1=u_4+u_5+u_6\eqno{(4-2-2a.3)}$$
Then, Together with (4-2.2a) we immediatly get
$$\begin{array}{l}
\check m_1(\eta)+\check m_4(\eta)=0\\
2\check m_1(\eta)=\check m_5(\eta)+\check m_6(\eta)
\end{array}
\eqno{(4-2-2a.4)}$$
Substituting it into the fifth equation of (2.10c) and its counterpart, we get
$$2\check m_1u_7(u_1-u_6)=u_1\frac{\partial u_7}{\partial \eta}-u_4\check m_7=u_1\check m_7-u_4\frac{\partial u_7}{\partial \eta}$$
Using the initial conditions (2.8) there is
$$\frac{\partial u_7}{\partial \eta}=\check m_7(\eta)\eqno{(4-2-2a.5)}$$
So, combining (4-2-2a.2), (4-2-2a,5) and the initial conditions(2.8), we can write down 
$$u_7=\alpha u+F(\xi)-F(\eta)\eqno{(4-2-2a.6)}$$
Substituting all the results into the sixth equation of (2.10a) as well as its counterpart, (4-2.2a) and (4-2-2c), we have
$$(u_1^2-u_5^2)\alpha -2m_1(\eta)u_7-\alpha =0\eqno{(4-2-2a.7a)}$$
$$(u_4^2-u_6^2)\alpha +2m_1(\eta)u_7-\alpha =0\eqno{(4-2-2a.7b)}$$
$$u_1u_4+u_5u_6=1\eqno{(4-2-2a.7c)}$$
$$(u_1u_6+u_4u_5)\alpha =2m_1(\eta)u_7\eqno{(4-2-2a.7d)}$$
Using the symmetric operation, we aslo have
$$(u_1^2-u_6^2)\alpha -2m_1(\xi)u_7-\alpha =0\eqno{(4-2-2a.8a)}$$
$$(u_4^2-u_5^2)\alpha +2m_1(\xi)u_7-\alpha =0\eqno{(4-2-2a.8b)}$$
$$u_1u_4+u_5u_6=1\eqno{(4-2-2a.8c)}$$
$$(u_1u_5+u_4u_6)\alpha =2m_1(\xi)u_7\eqno{(4-2-2a.8d)}$$
From (4-2-2a.7) we have
$$u_4(\alpha +2m_1(\eta)u_7)=\alpha u_1-2m_1(\eta)u_5u_7\eqno{(4-2-2a.9a)}$$
$$u_5(2m_1(\eta)u_7-\alpha )=\alpha u_6-2m_1(\eta)u_4u_7\eqno{(4-2-2a.9b)}$$
From (4-2-2a.8) we also get
$$u_1(\alpha -2m_1(\xi)u_7)=\alpha u_4-2m_1(\xi)u_5u_7\eqno{(4-2-2a.9c)}$$
$$u_6(2m_1(\xi)u_7-\alpha )=\alpha u_5-2m_1(\xi)u_4u_7\eqno{(4-2-2a.9b)}$$
Solving the system of equation (4-2-2a.9), we get
$$\frac{u_4}{u_1}=\frac{\alpha (m_1(\xi)+m_1(\eta))-2m_1(\xi)m_1(\eta)u_7}{\alpha (m_1(\xi)+m_1(\eta))+2m_1(\xi)m_1(\eta)u_7}\eqno{(4-2-2a.10a)}$$
$$\frac{u_6}{u_5}=\frac{-\alpha (m_1(\xi)-m_1(\eta))+2m_1(\xi)m_1(\eta)u_7}{\alpha (m_1(\xi)-m_1(\eta))+2m_1(\xi)m_1(\eta)u_7}\eqno{(4-2-2a.10b)}$$
We denote here
$$\begin{array}{l}
H_1=\alpha (m_1(\xi)+m_1(\eta))+2m_1(\xi)m_1(\eta)u_7\\[4mm]
H_4=\alpha (m_1(\xi)+m_1(\eta))-2m_1(\xi)m_1(\eta)u_7\\[4mm]
H_5=\alpha (m_1(\xi)-m_1(\eta))+2m_1(\xi)m_1(\eta)u_7\\[4mm]
H_6=-\alpha (m_1(\xi)-m_1(\eta))+2m_1(\xi)m_1(\eta)u_7
\end{array}$$
Then we can assum
$$\begin{array}{ll}
u_1=H_1X&u_4=H_4X\\[4mm]
u_5=H_5Y&u_6=H_6Y
\end{array}
\eqno{(4-2-2a.11)}$$
On the other hand, from (4-2-2a.7a) and (4-2-2a.8b), we have
$$\alpha (u_1^2-u_4^2)=2u_7(m_1(\xi)+m_1(\eta))$$
Similarly, from (4-2-2a.7b) and (4-2-2a.8b), we also have
$$\alpha (u_5^2-u_6^2)=2u_7(m_1(\xi)-m_1(\eta))$$
From them, we can get
$$X=Y=\frac{1}{2\alpha \sqrt{m_1(\xi)m_1(\eta)}}$$
So, we have obtained the solution of this case:
$$\begin{array}{l}
a_1(u,\xi,\eta)=\frac{1}{2\alpha G(\xi)G(\eta)}(\alpha (G(\xi)^2+G(\eta)^2)+2G(\xi)^2G(\eta)^2(\alpha u+F(\xi)-F(\eta)))\\[4mm]
a_2(u,\xi,\eta)=a_3(u,\xi,\eta)=1\\[4mm]
a_4(u,\xi,\eta)=\frac{1}{2\alpha G(\xi)G(\eta)}(\alpha (G(\xi)^2+G(\eta)^2)-2G(\xi)^2G(\eta)^2(\alpha u+F(\xi)-F(\eta)))\\[4mm]
a_5(u,\xi,\eta)=\frac{1}{2\alpha G(\xi)G(\eta)}(\alpha (G(\xi)^2-G(\eta)^2)+2G(\xi)^2G(\eta)^2(\alpha u+F(\xi)-F(\eta)))\\[4mm]
a_6(u,\xi,\eta)=\frac{1}{2\alpha G(\xi)G(\eta)}(-\alpha (G(\xi)^2-G(\eta)^2)+2G(\xi)^2G(\eta)^2(\alpha u+F(\xi)-F(\eta)))\\[4mm]
a_7(u,\xi,\eta)=\alpha u+F(\xi)-F(\eta)
\end{array}
\eqno{(4-2-2a.12)}$$

{\it (B). $\delta\not=0$ subcase}

Now, (4-2-2.1) changes to be
$$(u_7^{'})^2=\alpha ^2-\delta ^2u_7^2
\eqno{(4-2-2b.1)}$$
Thanks to remark 2. we also have
$$(\frac{\partial u_7}{\partial \eta})^2=\check m_7(\eta)^2-((\check m_5(\eta)+\check m_6(\eta))^2-4\check m_1(\eta)^2)u_7^2
\eqno{(4-2-2b.2)}$$
On the other hand, still because of remark 2, we have
$$(\check m_5(\eta)+\check m_6(\eta))^2-4\check m_1(\eta)^2=\frac{\check m_7(\eta)^2}{u_7^2}((u_1u_6+u_4u_5)^2-\frac{1}{4}(u_1^2+u_6^2-u_4^2-u_5^2))
=\frac{\delta ^2}{\alpha ^2}\check m_7(\eta)^2$$
So, (4-2-2b.2) can be written as
$$(\frac{\partial u_7}{\partial \eta})^2=\check m_7(\eta)^2-\frac{\delta ^2}{\alpha ^2}\check m_7(\eta)^2u_7^2
\eqno{(4-2-2b.3)}$$
Together with (4-2-2b.2) and the initial conditions (2.8), we have
$$a_7(u,\xi,\eta)=\frac{\alpha }{\delta }sin(\delta u+F(\xi)-F(\eta))
\eqno{(4-2-2b.4)}$$
where $F(\xi)$ is an arbitary function of coloured parameter.

Substituting all the results into the sixth equation of (2.10a) as well as its counterpart, (4-2.2a) and (4-2-2c), we have
$$\delta cos(\delta u+F(\xi)-F(\eta))+2m_1(\eta)sin(\delta u+F(\xi)
-F(\eta))-\delta (u_1^2-u_5^2)=0\eqno{(4-2-2b.5a)}$$
$$\delta cos(\delta u+F(\xi)-F(\eta))-2m_1(\eta)sin(\delta u+F(\xi)
-F(\eta))-\delta (u_4^2-u_6^2)=0\eqno{(4-2-2b.5b)}$$
$$u_1u_4+u_5u_6=1\eqno{(4-2-2b.5c)}$$
$$\delta (u_1u_6+u_4u_5)-2m_5(\eta)sin(\delta u+F(\xi)-F(\eta))=0\eqno{(4-2-2b.5d)}$$
Using the symmetric operation, we can also have
$$\delta cos(\delta u+F(\xi)-F(\eta))+2m_1(\xi)sin(\delta u+F(\xi)
-F(\eta))-\delta (u_1^2-u_6^2)=0\eqno{(4-2-2b.6a)}$$
$$\delta cos(\delta u+F(\xi)-F(\eta))-2m_1(\xi)sin(\delta u+F(\xi)
-F(\eta))-\delta (u_4^2-u_5^2)=0\eqno{(4-2-2b.6b)}$$
$$u_1u_4+u_5u_6=1\eqno{(4-2-2b.6c)}$$
$$\\delta (u_1u_5+u_4u_6)-2m_5(\xi)sin(\delta u+F(\xi)-F(\eta))=0\eqno{(4-2-2b.6d)}$$
As for the case of $\delta=0$, when we do the same procedures, and at last we can get 
$$\begin{array}{l}
\displaystyle\frac{u_4}{u_1}=\displaystyle\frac{\delta m_5(\xi)+\delta m_5(\eta)cos(\delta u+F(\xi)-F(\eta))-2m_1(\xi)m_5(\eta)sin(\delta u+F(\xi)-F(\eta))}{\delta m_5(\eta)+\delta m_5(\xi)cos(\delta u+F(\xi)-F(\eta))+2m_5(\xi)m_1(\eta)sin(\delta u+F(\xi)-F(\eta))}\\[4mm]
\displaystyle\frac{u_6}{u_5}=\displaystyle\frac{-\delta m_5(\eta)+\delta m_5(\xi)cos(\delta u+F(\xi)-F(\eta))-2m_5(\xi)m_1(\eta)sin(\delta u+F(\xi)-F(\eta))}{-\delta m_5(\xi)+\delta m_5(\eta)cos(\delta u+F(\xi)-F(\eta))-2m_1(\xi)m_5(\eta)sin(\delta u+F(\xi)-F(\eta))}
\end{array}
\eqno{(4-2-2b.7)}$$
Similarly we set here
$$\begin{array}{l}
H_1=\delta m_5(\eta)+\delta m_5(\xi)cos(\delta u+F(\xi)-F(\eta))+2m_5(\xi)m_1(\eta)sin(\delta u+F(\xi)-F(\eta))\\[4mm]
H_4=\delta m_5(\xi)+\delta m_5(\eta)cos(\delta u+F(\xi)-F(\eta))-2m_1(\xi)m_5(\eta)sin(\delta u+F(\xi)-F(\eta))\\[4mm]
H_5=-\delta m_5(\xi)+\delta m_5(\eta)cos(\delta u+F(\xi)-F(\eta))-2m_1(\xi)m_5(\eta)sin(\delta u+F(\xi)-F(\eta))\\[4mm]
H_6=-\delta m_5(\eta)+\delta m_5(\xi)cos(\delta u+F(\xi)-F(\eta))-2m_5(\xi)m_1(\eta)sin(\delta u+F(\xi)-F(\eta))
\end{array}
\eqno{(4-2-2b.8)}$$
We also assum
$$\begin{array}{ll}
u_1=H_1X&u_4=H_4X\\[4mm]
u_5=H_5Y&u_6=H_6Y
\end{array}
\eqno{(4-2-2b.9)}$$
Then from (4-2-2b.5) and (4-2-2b.6) we have
$$\begin{array}{l}
X=\sqrt{\displaystyle\frac{2(m_1(\xi)+m_1(\eta))sin(\delta u+F(\xi)-F(\eta))}{\delta (H_1^2-H_4^2)}}\\[4mm]
Y=\sqrt{\displaystyle\frac{2(m_1(\xi)-m_1(\eta))sin(\delta u+F(\xi)-F(\eta))}{\delta (H_5^2-H_6^2)}}
\end{array}
\eqno{(4-2-2b.10)}$$
So, we at last write down the solution of this case:
$$\begin{array}{l}
a_1(u,\xi,\eta)=X(\delta G(\eta)+\delta G(\xi)cos(\delta u+F(\xi)-F(\eta))+2G(\xi)H(\eta)sin(\delta u+F(\xi)-F(\eta)))\\[4mm]
a_2(u,\xi,\eta)=a_3(u,\xi,\eta)=1\\[4mm]
a_4(u,\xi,\eta)=X(\delta G(\xi)+\delta G(\eta)cos(\delta u+F(\xi)-F(\eta))-2G(\eta)H(\eta)sin(\delta u+F(\xi)-F(\eta)))\\[4mm]
a_5(u,\xi,\eta)=Y(-\delta G(\xi)+\delta G(\eta)cos(\delta u+F(\xi)-F(\eta))-2G(\eta)H(\xi)sin(\delta u+F(\xi)-F(\eta)))\\[4mm]
a_6(u,\xi,\eta)=Y(-\delta G(\eta)+\delta G(\xi)cos(\delta u+F(\xi)-F(\eta))-2G(\xi)H(\eta)sin(\delta u+F(\xi)-F(\eta)))\\[4mm]
a_7(u,\xi,\eta)=\displaystyle\frac{\alpha }{\delta }sin(\delta u+F(\xi)-F(\eta))
\end{array}
\eqno{(4-2-2b.11)}$$
where $F(\xi), G(\xi)$ and $H(\xi)$ are arbitary functions of coloured parameters and $X,Y$ are definded by equation (4-2-2b.10). By letting
$\delta \rightarrow 0$ we can see here that the solution above degenerate to
(4-2-2a.12).

If we let $\xi=\eta$ in the upper equations, from (4-2-2b.7) we can see that
$$ u_5=u_6$$
There is also a shortly expression for $X$ from (4-2-2b.10)
$$ X=\displaystyle\frac{1}{2\delta m_5 cos(\delta u/2)}$$
and we can easily get the corresponding solution of the YBE without coloured parameters
$$\begin{array}{l}
a_1(u)=\displaystyle\frac{\delta cos(\frac{\delta u}{2})+2 m_1 sin(\frac{\delta u}{2})}{\delta}\\[4mm]
a_4(u)=\displaystyle\frac{\delta cos(\frac{\delta u}{2})-2 m_1 sin(\frac{\delta u}{2})}{\delta}\\[4mm]
a_5(u)=a_6(u)=\displaystyle\frac{2 m_5 sin(\frac{\delta u}{2})}{\delta}\\[4mm]
a_7(u)=\frac{\alpha}{\delta}sin(\delta u)
\end{array}
$$
which is the same as the second solution of YBE without coloured parameters obtained in [25].

\section{General solutions}

In this paper, we have given five basic solutions and two degerate solutions of equation (2.1) and also classify them into two types.These seven solutions i.e. (3-2-1.8), (3-2-1.16), (4-1.16a), (4-1.16b), (4-2-1.6), (4-2-2a.12) and (4-2-2b.10) together with the five solution transformations {\bf A-E} will give all seven-vertex-type solutions of coloured Yang-Baxter equation (2.1) and the general solutions can also be classified into two types. The first are Baxter-type solutions which can be obtained from the basic Bxter-type solution via some solution transformations. The second are free-fermion-type which can be obtained via the basic free-fermion-type solutions via some solution transformations. Furthermore, we have showed that three of the five basic solutions can be degenerated into the solutions of non-coloured YBE obtained in [25]. But the other two solutions   can also be degenerated into pure spectral case which has not correspondence in [25].

According to the standard model given by Baxter, for a given R matrix the spin-chain Hamiltonian is generally of the following form:
$$H=\sum _{j=1}^N(J_x\sigma _j^x\sigma _{j+1}^x+J_y\sigma _j^y\sigma _{j+1}^y+J_z\sigma _j^z\sigma _{j+1}^z+\frac{1}{2}(\sigma _j^z+\sigma {j+1}^z))$$
where $\sigma ^x, \sigma ^y$ and $\sigma ^z$ are Pauli matrices and the coupling constant are
$$\begin{array}{ll}
J_x=\frac{1}{4}(m_5+m_6+m_7)&J_y=\frac{1}{4}(m_5+m_6-m_7)\\
J_z=\frac{1}{4}(m_1-m_3+m_4-m_2)&h=\frac{1}{4}(m_1-m_3-m_4+m_2)
\end{array}$$

From [22,23], we can know that in all the solutions of six-vertex and eight-vertex cases the Hamiltonian coefficients obey
$$\begin{array}{ll}
m_1^2=m_4^2&m_5^2=m_6^2
\end{array}$$
and in this paper, we have proved that in the solutions (3-2-1.16), (4-1.16a), (4-1.16b), (4-2-1.6), (4-2-2a.12) and (4-2-2b.10)   the relations above are also sound except in the solution   (3-2-1.8). What makes this difference remains to be investigated.

\vskip 7mm
\noindent{\bf ACKNOWLEDGMENTS}
\vskip 4mm
This work  supported by Climbing Up Project, NSCC, Natural 
Scientific Foundation of Chinese Academy of Sciences and 
Foundation of NSF.

\end{document}